\documentclass[11pt]{article}

\usepackage{amssymb}
\usepackage{citesort}
\usepackage{amsmath,mathrsfs}
\usepackage{url}

\def\la{\langle}
\def\ra{\rangle}
\def\pa{\partial}

\def\al{\alpha}

\def\ii{\textrm i}
\def\ee{\textrm e}
\def\vp{\varphi}

\newcommand{\db}{de$\,$Broglie}

\newcommand{\dbb}{de$\,$Broglie-Bohm}

\newcommand{\be}{\begin{equation}}
\newcommand{\en}{\end{equation}}
\newcommand{\bea}{\begin{eqnarray}}
\newcommand{\ena}{\end{eqnarray}}

\begin{document}

\vspace*{1.0cm}
\noindent
{\bf
{\large
\begin{center}
On the zig-zag pilot-wave approach for fermions
\end{center}
}
}

\vspace*{.5cm}
\begin{center}
Ward Struyve\\
Institute for Theoretical Physics, University of Leuven,\\
Celestijnenlaan 200D, B--3001 Leuven, Belgium.\\
E--mail: Ward.Struyve@fys.kuleuven.be.
\end{center}

\begin{abstract}
\noindent
We consider a pilot-wave approach for the Dirac theory that was recently proposed by Colin and Wiseman. In this approach, the particles perform a zig-zag motion, due to stochastic jumps of their velocity. We respectively discuss the one-particle theory, the many-particle theory and possible extensions to quantum field theory. We also discuss the non-relativistic limit of the one-particle theory. For a single particle, the motion is always luminal, a feature that persists in the non-relativistic limit. For more than one particle the motion is in general subluminal. 
\end{abstract}

\renewcommand{\baselinestretch}{1.1}
\bibliographystyle{unsrt}
\bibliographystyle{plain}

\section{Introduction} 
Pilot-wave theories form alternatives to standard quantum theory that are not plagued by the conceptual difficulties, such as the measurement problem, that plague the latter~\cite{bohm93,holland93b,durr09}. In the non-relativistic pilot-wave theory of \db\ and Bohm, systems are described by actual point particles which move in physical space under the guidance of the system's wave function. This theory has been extended to relativistic quantum theory and quantum field theory (for the latter see e.g.~\cite{durr05a,struyve10,struyve11}). In particular, for the quantized Dirac field, there exists a pilot-wave approach with positions for particles and anti-particles~\cite{durr05a} and a Dirac sea approach~\cite{colin07} (see~\cite{struyve11} for a comparison).{\footnote{These pilot-wave models introduce positions instead of field configurations. While field configurations can be straightforwardly introduced for bosons, this seems much harder for fermions~\cite{struyve10,struyve11}.}}  The former is stochastic, due to the possibility of particle creation and annihilation, while the latter is deterministic.

Recently, Colin and Wiseman have considered a novel pilot-wave approach for the quantized Dirac field~\cite{colin11}. This approach starts from a reformulation of the (massive) Dirac theory in terms of a pair of massless Weyl fields of opposite chirality. The Dirac mass then yields a coupling of these Weyl fields. As such, they were led to introduce actual positions for left- and right-handed Weyl particles, which could be done by adopting the particle--anti-particle picture or the Dirac sea picture. In the latter case, the chirality of the particles may change stochastically, causing a discontinuity in their velocity. This results in a zig-zag motion. In the special case of a single particle, the motion is always luminal. In the case of more than one particle, the motion is in general subluminal.

Similar zig-zag trajectories have been considered before in different contexts. For example, Feynman formulated a path integral approach for the Dirac equation in 2-dimensional space-time (called the {\em Feynman checkerboard}), in terms of paths of particles that move back and forth at the speed of light~\cite[pp.\ 34-36]{feynman65},~\cite{schweber86} (see~\cite{cartier06} for further developments of this model). Recently, also Penrose considered such zig-zag paths as Feynman graphs for Dirac fermions~\cite{penrose04}. 

Colin and Wiseman were motivated to study these alternative approaches by the fact that according to the standard model of particle physics, the fermions are fundamentally massless and acquire an effective mass only through the interaction with the Higgs field. Hence, in a fundamental pilot-wave approach, actual positions should be introduced for the massless particles and one way to do this is by introducing positions for Weyl particles. Not only could this be done by adopting the particle--anti-particle picture or the Dirac sea picture, but one can also further choose whether or not to distinguish particles of different chiralities. That is, instead of introducing particles that are distinguishable by chirality, one may also choose to introduce indistinguishable Weyl particles (equivalently one could say that one is introducing positions for massless Dirac particles). In the case of the Dirac sea approach, this leads to a deterministic theory (which was favored by Colin and Wiseman). 

In the present work, we further explore Colin and Wiseman's zig-zag pilot-wave approach to the Dirac theory. In section \ref{singleparticle}, we start with considering the single particle Dirac equation. By decomposing the Dirac spinor into a left- and right-handed chiral component, a set of coupled Weyl equations is obtained, for a left- and right-handed Weyl spinor. This suggests a pilot-wave model according to which there is an actual particle that moves along a continuous trajectory at the speed of light, but with a direction of velocity that changes stochastically and either depends on the left or right-handed spinor. This model is extended to systems of many particles in section \ref{manyparticles}. In this case, the particles do not necessarily move at the speed of light. In section \ref{qft}, the possible extensions to quantum field theory are discussed. The different possibilities arise from the fact that, first of all, as mentioned before, one can adopt the usual particle--anti-particle picture or the Dirac sea picture. Secondly, one can regard the massive Dirac particles as fundamental or the Weyl particles. Of course, in a more fundamental theory, where positions should be introduced for massless particles, one should opt for the latter choice. The different possible pilot-wave models in terms of Weyl particles are summarized in section \ref{differentmodels}. Then, in section \ref{nonrelativistic} we discuss the non-relativistic limit of the single particle model. Finally, in section \ref{variations}, we consider possible pilot-wave models for respectively the Pauli equation and the non-relativistic Schr\"odinger equation that are constructed in a similar way. The latter models should not be taken as serious alternatives to the usual approaches, but instead serve to illustrate theoretical possibilities.

\section{Single-particle Dirac theory}\label{singleparticle}
Before presenting the zig-zag pilot-wave model, we first consider Bohm's original pilot-wave approach~\cite{bohm53c}, which is deterministic. For a single particle, the guiding wave is a Dirac spinor $\psi(x)=\psi({\bf x},t)$, which takes values in ${\mathbb C}^{4}$, and which satisfies the Dirac equation{\footnote{Throughout the paper we use natural units in which $\hbar=c=1$.}} 
\begin{equation}
\ii \gamma^\mu D_\mu  \psi - m \psi = 0 \,,
\label{1}
\end{equation}
where $A^\mu = (V,{\bf A})$ is an external electromagnetic potential and $D_\mu =\partial_\mu + \ii e A_\mu$ is the covariant derivative. In Bohm's approach, the possible world lines $x^\mu(s)$ of the particle, parametrized by $s$, satisfy the guidance equation
\begin{equation}
\frac{d x^\mu }{d s} = j^\mu_D = {\bar \psi} \gamma^\mu \psi \,,
\label{2}
\end{equation}
where ${\bar \psi} = \psi^\dagger \gamma^0$ and $j^\mu_D$ is the usual Dirac current, which is conserved, i.e.\ 
\be
\pa_\mu j^\mu_D=0 \,. 
\label{2.1}
\en
(Note that multiplication of the current by a function $a(x)$ merely changes the parametrization of the world-lines and not their image in Minkowski space-time~\cite{berndl96a}). With respect to a particular reference frame, the path can be expressed as ${\bf x}(t)$. The guidance equation then reads
\begin{equation} 
\frac{d {\bf x}}{d t} = {\bf v}_D = \frac{{\bf j}_D}{j^0_D} = \frac{\psi^\dagger {\boldsymbol \alpha}  \psi}{\psi^\dagger \psi} \,.
\label{3}
\end{equation}
Since $j^\mu_D j_{D\mu} \ge 0$, $|{\bf v}_D| \le 1$, so that the velocity never exceeds the speed of light \cite{holland93b}. Moreover, for generic wave functions, the probability that the speed of light is reached is zero \cite{tausk10}.

For an ensemble of systems all with the same Dirac spinor $\psi$, Bohm's pilot-wave approach reproduces the predictions of standard quantum theory, given that the particle distribution is given by $\psi^\dagger \psi$. This is called the quantum equilibrium distribution. The guidance equation implies that if the distribution is given by $\psi^\dagger({\bf x},t_0) \psi({\bf x},t_0)$ at a certain time $t_0$, it will be given by $\psi^\dagger({\bf x},t) \psi({\bf x},t)$ at other times $t$, a property called {\em equivariance}~\cite{durr09}. This follows from the fact that an arbitrary distribution $\rho({\bf x},t)$ that is transported along the trajectories satisfies the continuity equation
\be
\pa_t \rho + {\boldsymbol \nabla} \cdot \left( {\bf v}_D \rho \right) = 0 
\en
and that the density $\psi^\dagger \psi$ satisfies this equation because of \eqref{2.1}.
 
The zig-zag pilot-wave model forms an alternative approach. It can be obtained as follows.{\footnote{Colin and Wiseman~\cite{colin11} obtained this model in the context of quantum field theory, adopting the Dirac sea picture. We will turn to quantum field theory in section \ref{qft}.}} Using the chiral projection operators 
\be
P_R= (1 + \gamma_5)/2 \,, \quad P_L= (1 - \gamma_5)/2 \,,
\label{4}
\en 
the Dirac spinor can be decomposed into a right- and left-handed chiral component: $\psi_R = P_R\psi$, $\psi_L = P_L\psi$.{\footnote{The decomposition of the spinor corresponds to the fact that the Dirac spinor transforms according to a reducible representation of the Lorentz group, given by ${\mathcal D}^{\left(\frac{1}{2}, 0\right)} \oplus {\mathcal D}^{\left(0, \frac{1}{2}\right)}$~\cite{peskin95}. The left- and right-handed spinor respectively transform according to the ${\mathcal D}^{\left(\frac{1}{2}, 0\right)}$ and ${\mathcal D}^{\left(0, \frac{1}{2}\right)}$ representation. A parity transformation transforms one into the other.}} Using this decomposition, the Dirac equation reduces to
\be
\ii \gamma^\mu D_\mu  \psi_R = m \psi_L \,, \qquad  \ii \gamma^\mu D_\mu  \psi_L = m \psi_R \,.
\label{5}
\en
The Dirac current decomposes as $j^\mu_D =   j^\mu_R + j^\mu_L$, where
\be
j^\mu_R = {\bar \psi}_R \gamma^\mu \psi_R  \,, \qquad j^\mu_L = {\bar \psi}_L \gamma^\mu \psi_L \,.
\label{6}
\en
These currents are light-like, i.e.\ $j^\mu_Rj_{R\mu} = j^\mu_Lj_{L\mu} =0$~\cite{peskin95,colin11}. They are not conserved, but instead satisfy
\begin{equation}
\pa_\mu j^\mu_R = F \,, \qquad \pa_\mu j^\mu_L = -F \,,
\label{7}
\end{equation}
where
\begin{equation}
F = 2m {\textrm{Im}} \left(\psi^\dagger_R  \gamma^0 \psi_L \right) \,.
\label{8}
\end{equation}
The equations \eqref{7} can be written as
\begin{align}
\pa_t \rho_R + {\boldsymbol \nabla} \cdot \left( {\bf v}_R \rho_R \right) &= t_{LR} \rho_L - t_{RL} \rho_R\,, \nonumber\\
\pa_t \rho_L + {\boldsymbol \nabla} \cdot \left( {\bf v}_L \rho_L \right) &= t_{RL} \rho_R - t_{LR} \rho_L \,,
\label{8.1}
\end{align}
where 
\begin{equation}
\rho_R = j^0_R \,,\qquad  \rho_L = j^0_L \,, \qquad {\bf v}_R = \frac{{\bf j}_R}{j^0_R} = \frac{\psi^\dagger_R {\boldsymbol \alpha}  \psi_R}{\psi^\dagger_R \psi_R} \,, \qquad {\bf v}_L = \frac{{\bf j}_L}{j^0_L} = \frac{\psi^\dagger_L {\boldsymbol \alpha}  \psi_L}{\psi^\dagger_L \psi_L}\,,
\label{9}
\end{equation}
\begin{equation}
t_{LR} = \frac{F^+}{j^0_L} = 2m \frac{ {\textrm{Im}}\left(\psi^\dagger_R \gamma^0  \psi_L \right)^+}{\psi^\dagger_L  \psi_L}  \,, \qquad t_{RL} = \frac{(-F)^+}{j^0_R} = 2m \frac{ {\textrm{Im}} \left(\psi^\dagger_L \gamma^0 \psi_R \right)^+ }{\psi^\dagger_R  \psi_R}\,,
\label{10}
\end{equation}
with $F^+=\max(F,0)$. 

The equations \eqref{8.1} now suggest a pilot-wave model where the particles move along continuous trajectories, with a velocity field that stochastically jumps between ${\bf v}_R$ or ${\bf v}_L$, with space-time dependent jump rates $t_{LR}$ and $t_{RL}$ to jump  respectively from ${\bf v}_L$ to ${\bf v}_R$ and vice versa.{\footnote{Note that Colin and Wiseman also labeled the positions with the chirality. We will not do this here.}} Since the currents $j^\mu_R$ and $j^\mu_L$ are light-like, $|{\bf v}_R| =|{\bf v}_L| = 1$ and the particles always move at the speed of light (a feature that in general does not hold any longer in the many-particle case~\cite{colin11}). 

For an ensemble of particles all guided by the same Dirac spinor $\psi$, the velocity phase space distribution 
\begin{equation}
\rho({\bf x},{\bf v},t) =  \rho_R({\bf x},t) \delta({\bf v} - {\bf v}_R({\bf x},t)) + \rho_L({\bf x},t) \delta({\bf v} - {\bf v}_L({\bf x},t)) 
\label{11}
\end{equation}
plays the role of equilibrium distribution. According to this distribution, the probabilities that a particle is in $d^3x$ around ${\bf x}$, with velocity respectively ${\bf v}_R$ and ${\bf v}_L$, are given by $j^0_R({\bf x},t)d^3x$ and $j^0_L({\bf x},t)d^3x$. This distribution is equivariant and the dynamics was chosen in order to guarantee this (see~\cite{durr02,durr031,durr032,durr05a} for a general discussion). Namely, for an arbitrary distribution  
\begin{equation}
p({\bf x},{\bf v},t) =  p_R({\bf x},t) \delta({\bf v} - {\bf v}_R({\bf x},t)) + p_L({\bf x},t) \delta({\bf v} - {\bf v}_L({\bf x},t)) \,,
\label{10.1}
\end{equation}
the spatial distributions $p_R$ and $p_L$ satisfy the master equations:
\begin{align}
\pa_t p_R + {\boldsymbol \nabla} \cdot \left( {\bf v}_R p_R \right) &= t_{LR} p_L - t_{RL} p_R\,, \nonumber\\
\pa_t p_L + {\boldsymbol \nabla} \cdot \left( {\bf v}_L p_L \right) &= t_{RL} p_R - t_{LR} p_L \,.
\label{10.2}
\end{align}
Because of \eqref{8.1}, the distribution given by $p_R =\rho_R$ and $p_L =\rho_L$ satisfies this equation. 

The resulting position distribution is $\rho_R + \rho_L = j^0_R + j^0_L = j^0_D = \psi^\dagger \psi$. As such, in equilibrium, the predictions agree with those of Bohm's pilot-wave approach for the Dirac theory, and hence with the standard quantum predictions. (For non-equilibrium distributions, as studied for example in~\cite{valentini07}, the pilot-wave approaches yield different predictions.)

The theory can be formulated more compactly as follows. By introducing a configuration $({\bf x},c) \in {\mathbb R}^{3} \times \{R,L\}$, given by a position ${\bf x}$ and chirality $c$, the trajectories are determined by  
\be
\frac{d {\bf x}}{d t} = {\bf v}_c 
\en
and the chirality can jump from $c$ to the opposite chirality, denoted by $\pi c$, with jump rates $t_{c,  \pi c}$. The equilibrium distribution reads $\rho_c({\bf x},t)$.

A similar approach was proposed before by de Angelis {\em et al.}~\cite{deangelis86} for the case of one spatial dimension (who considered it as a generalization of Nelson's stochastic mechanics~\cite{nelson66,goldstein87}). In this case, the velocity fields read $v_R=1$ and $v_L=-1$ (i.e., plus or minus the speed of light). The proposed jump rates are actually different from those considered here (requiring equivariance of \eqref{11} does not uniquely determine the rates). The jump rates here are {\em minimal} in the sense explained in~\cite{durr05a}. 

In the following, it will sometimes be convenient to use an explicit representation for the $\gamma$-matrices. Here, we choose the Dirac-Pauli representation. In this representation, we have, with $\psi = ({\widetilde \varphi}, {\widetilde \chi})^T$, where ${\widetilde \varphi}$ and ${\widetilde \chi}$ are 2-spinors, that 
\be
\psi_R = \frac{1}{\sqrt{2}} \left( \begin{array}{c} \vp_R \\ \vp_R \end{array} \right) \,, \qquad \psi_L = \frac{1}{\sqrt{2}} \left( \begin{array}{c} \vp_L \\ -\vp_L \end{array} \right) \,,
\label{12}
\en
where $\vp_R = ({\widetilde \varphi} + {\widetilde \chi})/{\sqrt 2}$ and $\vp_L = ({\widetilde \varphi} - {\widetilde \chi})/{\sqrt 2}$ are Weyl spinors. As such, the wave equations \eqref{5} become
\begin{equation}
\ii \sigma^\mu D_\mu  \vp_R = m \vp_L \,, \qquad  \ii {\bar \sigma}^\mu D_\mu  \vp_L = m \vp_R \,,
\label{13}
\end{equation}
where $\sigma^\mu = ( 1, {\boldsymbol \sigma})$ and ${\bar \sigma}^\mu = (1, - {\boldsymbol \sigma})$. The mass yields a coupling between the left- and right-handed spinor. In the case of zero mass, these equations decouple and respectively yield the right- and left-handed Weyl equation.

We further have that
\be
j^\mu_R = \vp^\dagger_R \sigma^\mu \vp_R = (\vp^\dagger_R  \vp_R, \vp^\dagger_R {\boldsymbol \sigma} \vp_R)\,, \qquad j^\mu_L = \vp^\dagger_L {\bar \sigma}^\mu \vp_L = (\vp^\dagger_L  \vp_L, - \vp^\dagger_L {\boldsymbol \sigma} \vp_L)\,,
\label{14}
\en
\begin{equation}
F = 2m {\textrm{Im}} \left(\vp^\dagger_R  \vp_L \right) \,,
\label{15}
\end{equation}
so that the velocities and jump rates can respectively be written as
\begin{equation}
{\bf v}_R = \frac{{\bf j}_R}{j^0_R} = \frac{\vp^\dagger_R {\boldsymbol \sigma}  \vp_R}{\vp^\dagger_R \vp_R} \,, \qquad {\bf v}_L = \frac{{\bf j}_L}{j^0_L} = - \frac{\vp^\dagger_L {\boldsymbol \sigma}  \vp_L}{\vp^\dagger_L \vp_L}\,,
\label{16}
\end{equation}
\begin{equation}
t_{LR} = \frac{F^+}{j^0_L} = 2m \frac{ {\textrm{Im}}\left(\vp^\dagger_R  \vp_L \right)^+}{\vp^\dagger_L  \vp_L}  \,, \qquad t_{RL} = \frac{(-F)^+}{j^0_R} = 2m \frac{ {\textrm{Im}} \left(\vp^\dagger_L  \vp_R \right)^+ }{\vp^\dagger_R  \vp_R}\,.
\label{17}
\end{equation}

\section{Many-particle Dirac theory}\label{manyparticles}
The zig-zag pilot-wave model for a single particle can straightforwardly be extended to many-particle systems. The wave function $\psi=\psi({\bf x}_1, \dots,{\bf x}_n,t)$ now takes values in the $n$-particle spin space $({\mathbb C}^{4})^{\otimes n}$ and satisfies the $n$-particle Dirac equation
\be
\ii \pa_t \psi =  \sum^n_{i=1} \left[- \ii {\boldsymbol \alpha}_i \cdot ({\boldsymbol \nabla}_i -\ii e {\bf A}({\bf x}_i,t))+ e V({\bf x}_i,t) + m \beta_i    \right] \psi \,,
\label{30}
\en
where ${\boldsymbol \alpha}_i = 1 \otimes \dots \otimes 1 \otimes {\boldsymbol \alpha} \otimes 1 \otimes \dots 1$, with ${\boldsymbol \alpha}$ at the $i$th of the $n$ places in the product, and similarly for $\beta_i$. (For simplicity, we have assumed equal masses and charges for the particles.) 

We define the projection operators 
\be
P_c = P_{c_1} \otimes \dots \otimes P_{c_n} \,,
\label{30.1}
\en
where $c = (c_1,\dots,c_n)$, with $c_i=R,L$, which are obtained by taking the product of $n$ chiral projections operators. As such, $\psi = \sum_c \psi_c$, where $\psi_c = P_c \psi$ and the sum is over all possible values of the chiralities. Application of the projection operator $P_c$ to the Dirac equation yields
\be
\ii \pa_t \psi_c =  \sum^n_{i=1} \left[- \ii {\boldsymbol \alpha}_i \cdot ({\boldsymbol \nabla}_i -\ii e {\bf A}({\bf x}_i,t))  + e V({\bf x}_i,t) \right] \psi_c + m \sum^n_{i=1} \beta_i   \psi_{\pi_i c} \,,
\label{31}
\en
where $\pi_i c$ is obtained from $c$ by changing the $i$th index $c_i$ from $R$ to $L$ or vice versa. It follows that
\be
\pa_t \rho_c + \sum^n_{i=1} {\boldsymbol \nabla}_i \cdot ({\bf v}_{i,c} \rho_c) =  \sum^n_{i=1} F_{\pi_i c,c} \,,
\label{32}
\en
where
\be
\rho_c = \psi^\dagger_c \psi_c  \,, \quad {\bf v}_{i,c} = \frac{\psi^\dagger_c {\boldsymbol \alpha}_i \psi_c}{\psi^\dagger_c \psi_c} \,, \quad F_{\pi_i c,c} = 2m {\textrm{Im}} \left(\psi^\dagger_c  \beta_i \psi_{\pi_i c} \right) \,.
\label{33}
\en

We can now assume $n$ particles, with positions ${\bf x}_1,\dots,{\bf x}_n$, whose velocity field is given by $v_c= ({\bf v}_{1,c} , \dots,{\bf v}_{n,c})$ for a certain $c$, and jump rates 
\be
t_{c,\pi_i c} = \frac{F^+_{c,\pi_i c}}{\rho_c}
\label{34}
\en
for the velocity field $v_c$ to jump to $v_{\pi_i c}$. Note that while a jump will only change one of the chiralities, it will in general lead to a discontinuity in the velocity of all the particles due to entanglement of the wave function (in particular, this will be the case of identical particles for which the wave function is assumed to be completely anti-symmetric under simultaneous exchange of position and spinor index). The velocity will not exceed the speed of light, but unlike the single particle case, it will in general not equal the light speed~\cite{colin11}.

In quantum equilibrium, given by the distribution $\rho_c$, the position distribution is given by $\sum_c \rho_c = \psi^\dagger \psi$, so that this approach yields the same predictions as the usual pilot-wave approach to the many-particle Dirac theory (which was presented in~\cite{bohm93}).

In the Dirac-Pauli representation, we can introduce the many-particle wave function $\vp_c$ which takes values in $({\mathbb C}^{2})^{\otimes n}$ and which is defined as $\vp_{c,\alpha} = \sqrt{2^n} \psi_{c,\alpha}$, where $\al=(\al_1,\dots,\al_n)$, with $\al_i = 1,2$, are the spinor indices. As such the wave equations can be written as
\be
\ii \pa_t \vp_c=  \sum^n_{i=1} \left[- s(c_i) \ii {\boldsymbol \sigma}_i \cdot ({\boldsymbol \nabla}_i -\ii e {\bf A}({\bf x}_i,t))  + e V({\bf x}_i,t) \right] \vp_c + m \sum^n_{i=1} \vp_{\pi_i c} \,,
\label{35}
\en
where
\be
s(R)=1\,, \qquad s(L)=-1 \,.
\label{35.1}
\en
The equilibrium distribution, the velocity field and transition rates can respectively be written as
\be
\rho_c =  \vp^\dagger_c \vp_c \,, \quad {\bf v}_{i,c} = s(c_i) \frac{\vp^\dagger_c {\boldsymbol \sigma}_i \vp_c}{\vp^\dagger_c \vp_c} \,, \quad t_{c,\pi_i c} = 2m \frac{ {\textrm{Im}} \left(\vp^\dagger_{\pi_i c} \vp_c \right) }{\vp^\dagger_c \vp_c} \,.
\label{36}
\en

\section{Quantum field theory}\label{qft}
As we will see, there are a number of ways to extend this theory to the quantized Dirac field. First, one can choose to introduce positions for massive Dirac particles or Weyl particles. (Of course, in a more fundamental approach to the standard model where particles are considered massless, we should not consider the first option.) Second, one can choose to adopt the particle--anti-particle picture or the Dirac sea picture. We will briefly explore these different approaches here. We will see that in employing the Dirac sea picture there is no difference in the two approaches. On the other hand, there is a difference in employing the particle--anti-particle picture.

\subsection{General structure of a pilot-wave model with stochastic jumps}
We start with recalling the general formalism for developing a pilot-wave model for quantum field theory which includes stochastic jumps of the configuration~\cite{durr02,durr031,durr032,durr05a} (also called {\em Bell-type quantum field theories}). The starting point is the Schr\"odinger equation 
\be
\ii \frac{d |\Psi(t)\ra}{dt} = {\widehat H} |\Psi(t)\ra 
\label{49}
\en
for the state vector $|\Psi\ra$. For simplicity, we will only consider the free Dirac field, for which the Hamiltonian is 
\be
{\widehat H} = \int d^3x {\widehat \psi}^\dagger({\bf x}) (- \ii {\boldsymbol \alpha} \cdot {\boldsymbol \nabla}  + m \beta ) {\widehat \psi}({\bf x}) \,.
\label{50}
\en
The Dirac field operators satisfy the anti-commutation relations $\{{\widehat \psi}_\alpha({\bf x}),{\widehat \psi}^\dagger_{\alpha'}({\bf x}')\} = \delta_{\alpha \alpha'} \delta({\bf x} - {\bf x}')$ and the other fundamental anti-commutation relations are zero. When considering the particle--anti-particle picture, we will always use the normal ordered Hamiltonian, denoted by $:{\widehat H}:$. In this way the vacuum state always has zero energy. 

The Hamiltonian can also be written as 
\begin{multline}
{\widehat H} = \int d^3x  \Big( \ii {\widehat \vp}^\dagger_L({\bf x}){\boldsymbol \sigma} \cdot {\boldsymbol \nabla} {\widehat \vp}_L({\bf x}) - \ii{\widehat \vp}^\dagger_R({\bf x}){\boldsymbol \sigma} \cdot {\boldsymbol \nabla} {\widehat \vp}_R({\bf x})  \\
+ m\left[ {\widehat \vp}^\dagger_L({\bf x})  {\widehat \vp}_R({\bf x}) + {\widehat \vp}^\dagger_R({\bf x}) {\widehat \vp}_L({\bf x}) \right]\Big) \,,
\label{51}
\end{multline}
where the Weyl field operators ${\widehat \vp}_c$ and ${\widehat \vp}^\dagger_c$, $c=R,L$, are defined through the operator equivalent of \eqref{12}, i.e.
\be
{\widehat \psi}_R = \frac{1}{\sqrt{2}} \left( \begin{array}{c} {\widehat \vp}_R \\ {\widehat \vp}_R \end{array} \right) \,, \quad {\widehat \psi}_L = \frac{1}{\sqrt{2}} \left( \begin{array}{c} {\widehat \vp}_L \\ -{\widehat \vp}_L \end{array} \right) \,.
\label{51.1}
\en 
They satisfy the anti-commutation relations $\{ {\widehat \vp}_{c,\alpha}({\bf x}),{\widehat \vp}^\dagger_{c',\alpha'}({\bf x}')\} = \delta_{cc'}\delta_{\alpha \alpha'} \delta({\bf x} - {\bf x}')$. 

The corresponding pilot-wave model will describe actual point-particles. The number of particles may change over time due to particle creation and annihilation. In the case there is only one type of particles, the configuration space is given by $\cup_{n \in {\mathbb N}} \left( {\mathbb R}^{3n} \times \{R,L\}^n \right)$. An element $(x,c)$ of this configuration space represents a configuration $x=({\bf x}_1, \dots, {\bf x}_n)$ of positions and a collection of chiralities $c=(c_1,\dots,c_n)$. In the case there are different types of particles, the configuration space will be given by the Cartesian product of the single particle configuration spaces. In between jumps, the particles will follow continuous trajectories determined by one of $2^n$ velocity fields, labeled by the different chiralities. Two types of jumps may occur. First, there may be a jump of just the chirality, which leads to a jump of the velocity field of the particles. Second, there may be jumps due to particle creation and annihilation. 

The construction of the model requires the choice of a suitable POVM ${\widehat P}_c(dx)$. The probability that the system with state $|\Psi(t)\ra$ is localized in $dx$, with chirality $c$, at time $t$, is given by $\langle \Psi(t) |{\widehat P}_c(dx) | \Psi(t) \rangle = \rho_c(x,t)dx$. The POVM $\sum_c {\widehat P}_c(dx)$, where the sum is over all possible chiralities, should be an appropriate position POVM, such that the usual quantum predictions are obtained in quantum equilibrium. 

The choice of dynamics will now be such that it leaves the distribution $\rho_c(x)$ equivariant. It is found by considering the equation
\be
\pa_t \langle \Psi(t) |{\widehat P}_c(dx) | \Psi(t) \rangle + 2 {\textrm{Im}} \langle \Psi(t) |{\widehat H} {\widehat P}_c(dx)  | \Psi(t) \rangle = 0 \,,
\label{52}
\en  
which follows from the Schr\"odinger equation. The goal is to bring this equation into the form 
\be
\pa_t \rho_c(x,t) + \nabla \cdot \left[ v_c(x,t) \rho_c(x,t) \right] = \sum_{c'} \int dx' \left[ t_{c'c}(x',x,t) \rho_{c'}(x',t) - t_{cc'}(x,x',t) \rho_c(x,t)\right] \,,
\label{53}
\en 
where the sum is over all chiralities and the integral over all possible position configurations, and where
\be
 \nabla \cdot \left[ v_c(x,t) \rho_c(x,t) \right]  dx = 2 {\textrm{Im}} \langle \Psi (t)|{\widehat H}_1 {\widehat P}_c(dx)  | \Psi (t)\rangle \,,
\label{54}
\en 
\be
t_{cc'}(x,x',t) dx' = 2\frac{ \left[ {\textrm{Im}}\langle \Psi (t)| {\widehat P}_{c'}(dx') {\widehat H}_2 {\widehat P}_c(dx) | \Psi (t)\rangle \right]^+}{\langle \Psi(t) |{\widehat P}_c(dx) | \Psi(t) \rangle} \,,
\label{55}
\en  
and ${\widehat H} = {\widehat H}_1 + {\widehat H}_2$ corresponds to some natural decomposition of the Hamiltonian. The equation \eqref{53} then suggest a pilot-wave model with velocity fields $v_c(x,t)$ and jump rates $t_{cc'}(x,x',t) dx'$ for a configuration to jump from $x$ to an element $dx'$ around $x'$ and from $c$ to $c'$.

We may obtain a pilot-wave model with no distinction between different chiralities, by starting from the position POVM $\sum_c {\widehat P}_c(dx)$, where the sum is over all possible chiralities. (See~\cite{durr032,durr05a,goldstein05b} for a general discussion on how to construct a pilot-wave approach in terms of identical particles.)

\subsection{Massive Dirac particles}
\subsubsection{Particles and anti-particles}
Consider the usual plane waves
\be
u_s({\bf p}) \ee^{\ii {\bf p} \cdot {\bf x}} \,, \quad v_s({\bf p}) \ee^{- \ii {\bf p} \cdot {\bf x}} \,,
\label{56}
\en
which are eigenstates of the Dirac Hamiltonian $- \ii {\boldsymbol \alpha} \cdot {\boldsymbol \nabla}  + m \beta$ with eigenvalues respectively $E_p$ and $-E_p$, where $E_p = \sqrt{|{\bf p}|^2 + m^2}$, and with momentum respectively ${\bf p}$ and $-{\bf p}$ (see e.g.~\cite{mandl86}). The label $s=1,2$ denotes the helicity. The Dirac field operator can be expanded in terms of these:
\be
{\widehat \psi}({\bf x}) = \sum_s \int \frac{d^3 p}{\sqrt{(2\pi)^3}} \sqrt{\frac{m}{E_p}} \left[ {\widehat b}_s({\bf p}) u_s({\bf p})\ee^{\ii {\bf p} \cdot {\bf x}}  + {\widehat d}^\dagger_s({\bf p}) v_s({\bf p}) \ee^{- \ii {\bf p} \cdot {\bf x}} \right] \,.
\label{57}
\en
The operators ${\widehat b}_s({\bf p})$ and ${\widehat d}_s({\bf p})$ respectively annihilate a particle and an anti-particle with momentum ${\bf p}$ and helicity $s$. 

Writing ${\widehat \psi}({\bf x}) = {\widehat b}({\bf x}) + {\widehat d}^\dagger({\bf x})$, where ${\widehat b}$ and ${\widehat d}^\dagger$ are respectively the particle and the anti-particle part of the field operator ${\widehat \psi}$, the Hamiltonian reduces to: 
\begin{align}
{\widehat H} &= \sum_{\alpha,\beta} \int d^3x \left[{\widehat b}^\dagger_\alpha({\bf x}) + {\widehat d}_\alpha({\bf x})\right] (- \ii {\boldsymbol \alpha} \cdot {\boldsymbol \nabla}  + m \beta )_{\alpha \beta} \left[{\widehat b}_\beta({\bf x}) + {\widehat d}^\dagger_\beta({\bf x}) \right]\nonumber\\
 &= \sum_{\alpha,\beta}\int d^3x \left[ {\widehat b}^\dagger_\alpha({\bf x}) (- \ii {\boldsymbol \alpha} \cdot {\boldsymbol \nabla}  + m \beta )_{\alpha \beta} {\widehat b}_\beta({\bf x})  +  {\widehat d}_\alpha ({\bf x})(- \ii {\boldsymbol \alpha} \cdot {\boldsymbol \nabla}  + m \beta )_{\alpha \beta} {\widehat d}^\dagger_\beta({\bf x}) \right]\nonumber\\
&= {\widehat H}_{D,1} +  {\widehat H}_{D,2} \,,
\label{57.1}
\end{align}
where 
\begin{align}
{\widehat H}_{D,1} &= \sum_{\alpha,\beta}\int d^3x \left[ {\widehat b}^\dagger_\alpha({\bf x})(- \ii {\boldsymbol \alpha}_{\alpha \beta} \cdot {\boldsymbol \nabla} ) {\widehat b}_\beta({\bf x})  + {\widehat d}_\alpha ({\bf x})(- \ii {\boldsymbol \alpha}_{\alpha \beta} \cdot {\boldsymbol \nabla}   ) {\widehat d}^\dagger_\beta({\bf x})\right] \,, \nonumber\\
{\widehat H}_{D,2} &=  m \sum_{\alpha,\beta}\int d^3x \left[ {\widehat b}^\dagger_\alpha({\bf x})  \beta_{\alpha \beta} {\widehat b}_\beta({\bf x})  + {\widehat d}_\alpha ({\bf x})\beta_{\alpha \beta} {\widehat d}^\dagger_\beta({\bf x})\right]\,.
\label{57.2}
\end{align}
In the second line of \eqref{57.1}, we have used the fact that the cross-terms containing particle and anti-particle operators vanish (which essentially follows from the property $u^\dagger_s({\bf p})v_{s'}(-{\bf p}) =0$~\cite{mandl86}). This leads to a decoupling of the particle and anti-particle part in the Hamiltonian. In the Schr\"odinger equation, we use the normal ordered Hamiltonian $:{\widehat H}:$. 

Consider now the states
\be
|x,\alpha;n,{\bar n}\rangle_D =  \frac{1}{\sqrt{n!{\bar n}!}} {\widehat b}^\dagger_{\alpha_1}({\bf x}_1)  \dots  {\widehat b}^\dagger_{\alpha_n}({\bf x}_n)  {\widehat d}^\dagger_{\alpha_{n+1}}({\bf x}_{n+1})  \dots   {\widehat d}^\dagger_{\alpha_{n+{\bar n}}}({\bf x}_{n+{\bar n}}) |0\ra_D \,,
\label{58}
\en
where $n,{\bar n}\in{\mathbb N}$, $x=({\bf x}_1, \dots,{\bf x}_{n+{\bar n}})$, $\alpha=(\alpha_1,\dots,\alpha_{n+{\bar n}})$, with $\alpha_i=1,\dots,4$, are the spinor indices, and $|0\ra_D$ is the state which contains no particles or anti-particles, i.e., ${\widehat b}_\alpha({\bf x})|0\ra_D= {\widehat d}_\alpha({\bf x})|0\ra_D = 0$ for all ${\bf x}$ and spinor indices $\alpha$. These states correspond to $n$ particles and ${\bar n}$ anti-particles, and span the Hilbert space. They are anti-symmetric under simultaneous exchange of positions and spin indices corresponding either to particles or anti-particles. In~\cite{durr05a}, a pilot-wave model was constructed starting from the position POVM $P^{(n,{\bar n})}_{D}(dx) = \sum_\alpha |x,\alpha;n,{\bar n} \ra_D {}_D\la x,\alpha;n,{\bar n}| dx$. In this model, there are positions for particles and anti-particles. In the absence of interactions, there are no jumps. In the case of for example electromagnetic interaction, there are jumps corresponding to particle creation and annihilation. 

Here we consider a different POVM. Let us therefore first introduce the states
\be
|x,c,\alpha;n,{\bar n}\rangle_D  = \sum_\beta \left( P_c \right)_{\alpha,\beta} |x,\beta;n,{\bar n}\rangle_D  \,,
\label{59}
\en
where $c=(c_1,\dots,c_{n+{\bar n}})$ and $P_c$ is the projection operator defined in \eqref{30.1}. For a state $| \Psi(t)\ra$, the Schr\"odinger equation \eqref{49} (with a normal ordered Hamiltonian) implies that the expansion coefficients $\psi^{(n,{\bar n})}_{c,\alpha}(x,t) = {}_D\la x,c,\alpha;n,{\bar n} | \Psi(t)\ra$ satisfy the many-particle wave equation \eqref{31} for vanishing electromagnetic potentials.

We can now define the POVM $P^{(n,{\bar n})}_{D,c}(dx) = \sum_\alpha |x,c,\alpha;n,{\bar n} \ra_D {}_D\la x,c,\alpha;n,{\bar n}| dx$. In the corresponding pilot-wave model, the velocity field corresponding to $:{\widehat H}_{D,1}:$ is the one given in \eqref{33}, evaluated for the wave function $\psi^{(n,{\bar n})}_{c,\alpha}(x,t)$, and the jump rates, which are derived from $:{\widehat H}_{D,2}:$, are
\be
t^{(n,{\bar n})}_{cc'}(x,x',t) = \delta_{\pi_i c,c'}\delta(x-x') 2m \frac{\left[ {\textrm{Im}} \left(\psi^{(n,{\bar n}) \dagger}_{c'} (x,t) \beta_i \psi^{(n,{\bar n})}_c  (x,t)\right) \right]^+}{\psi^{(n,{\bar n}) \dagger}_c (x,t) \psi^{(n,{\bar n})}_c (x,t)}\,.
\label{60}
\en
Because of the delta-function $\delta(x-x')$, there are no jumps of the position configuration. In particular, there is no particle creation or annihilation (even though the wave function need not be a particle number eigenstate). There are only jumps of the velocity with rates that are the same as those introduced for the many-particle Dirac theory. As such, this pilot-wave model reduces to the one for the many-particle Dirac theory. This is of course an artifact of the free theory. In the case of for example electromagnetic interaction, particle--anti-particle pair creation will be possible, which will be reflected in the jump rates.

In this pilot-wave approach, particles and anti-particles were introduced as different particle species. One could also construct a model in which they are indistinguishable. Such a model could be obtained by starting from the alternative POVM which is obtained from $P^{(n,{\bar n})}_c(dx)$ by summing over $n$ and ${\bar n}$, keeping $n+{\bar n}$ fixed~\cite{durr032,durr05a,goldstein05b}. 

Note that in order to have a pilot-wave model which does not distinguish particles of different chirality, one could start from the alternative POVM obtained by summing $P^{(n,{\bar n})}_{D,c}(dx)$ over all the possible chiralities $c$. The resulting position POVM is just $P^{(n,{\bar n})}_{D}(dx)$, which was considered in~\cite{durr05a}.

\subsubsection{Dirac sea}\label{dirac-diracsea}
Instead of using the particle--anti-particle picture, one can also adopt the Dirac sea picture. The starting point is the field expansion 
\be
{\widehat \psi}({\bf x}) = \sum_s \int \frac{d^3 p}{\sqrt{(2\pi)^3}} \sqrt{\frac{m}{E_p}} \left[ {\widehat b}_s({\bf p}) u_s({\bf p})\ee^{\ii {\bf p} \cdot {\bf x}}  + {\widehat {\widetilde b}}_s({\bf p}) v_s(-{\bf p}) \ee^{ \ii {\bf p} \cdot {\bf x}} \right] \,,
\label{61}
\en
where ${\widehat {\widetilde b}}_s({\bf p})={\widehat d}^\dagger_s(-{\bf p})$ is the annihilation operator for a particle of negative energy $E_p$ and momentum ${\bf p}$. 

The states
\be
|x,\alpha\rangle_D = \frac{1}{\sqrt{n!}}{\widehat \psi}^\dagger_{\alpha_1}({\bf x}_1)  \dots  {\widehat \psi}^\dagger_{\alpha_n}({\bf x}_n)  |{\widetilde 0}\ra_D 
\label{61.1}
\en
can be introduced, where the state $|{\widetilde 0}\ra_D$ is now the state that does not contain particles of positive or negative energy, i.e., ${\widehat \psi}_\alpha({\bf x})|{\widetilde 0}\ra_D=0$ for all ${\bf x}$ and spinor indices $\alpha$. They are anti-symmetric under simultaneous exchange of positions and spin indices and span the Hilbert space. The state $|0\ra_D$ can be obtained by applying all the creation operators of negative energy particles to $|{\widetilde 0}\ra_D$.

In~\cite{colin07}, the position POVM $P_D(dx) = \sum_\alpha |x,\alpha\rangle_D {}_D\la x,\alpha| dx$ was considered. In the corresponding pilot-wave model, there is a fixed number of particles which move deterministically, even in the case of electromagnetic or other fermion number preserving types of interaction.

Consider now the states
\be
|x,c,\alpha\rangle_D = \sum_\beta \left( P_c \right)_{\alpha,\beta} |x,\beta\rangle_D \,.
\label{62}
\en
The expansion coefficients $\psi_{c,\alpha}(x,t) = {}_D\la x,c,\alpha | \Psi(t)\ra$ of a state $| \Psi(t)\ra$ satisfy the many-particle wave equation \eqref{31} as a consequence of the Schr\"odinger equation \eqref{49}. Starting from the position POVM $P_{D,c}(dx) = \sum_\alpha |x,c,\alpha \ra_D {}_D\la x,c,\alpha| dx$, we obtain the following pilot-wave model. The velocity field is determined by ${\widehat H}_{D,1}$ and corresponds to the one given in \eqref{33} for the wave function $\psi_{c,\alpha}(x,t)$ and the jump rates derived from ${\widehat H}_{D,2}$ are of the same form as in \eqref{60}. So again, the pilot-wave model is formally similar to the one for the many-particle theory. This is still the case when electromagnetic interactions are considered, unlike the zig-zag model obtained in the particle--anti-particle picture. Another difference with the latter model is the number of particles. For example, the case where there are no particles or anti-particles corresponds to a filled Dirac sea and hence to an infinite number of particles in the present model (employing regulators the actual number of particles could be made finite~\cite{colin07}).

\subsection{Weyl particles}
\subsubsection{Particles and anti-particles}
Instead of expanding the field operators in terms of eigenstates of the Dirac Hamiltonian, they can also be expanded in terms of eigenstates of the right- and left-chiral Weyl Hamiltonians $\mp \ii {\boldsymbol \sigma} \cdot {\boldsymbol \nabla}$~\cite{colin11}. By introducing the 2-component spinors $w_i({\bf p})$, $i=1,2$, with 
\be
\frac{{\boldsymbol \sigma} \cdot {\bf p}}{|{\bf p}|} w_i({\bf p}) = (-1)^{1+i} w_i({\bf p}) \,,
\label{63}
\en
we have that the plane waves 
\be
w_1({\bf p}) \ee^{\ii {\bf p} \cdot {\bf x}} \,, \quad w_2(-{\bf p}) \ee^{- \ii {\bf p} \cdot {\bf x}}
\label{64}
\en
are eigenstates of the right-chiral Weyl Hamiltonian with eigenvalues respectively $|{\bf p}|$ and $-|{\bf p}|$. Since they have momentum respectively ${\bf p}$ and $-{\bf p}$, the relations \eqref{63} imply that they have respectively right- and left-handed helicity, i.e.\ their momentum direction is respectively aligned and anti-aligned with their spin. (More generally, for positive energy plane waves, the chirality equals the helicity. For negative energy plane waves, they are opposite.) Similarly, we have eigenstates of the left-chiral Weyl Hamiltonian.

In terms of the plane wave expansion, the field operators read
\begin{align}
{\widehat \vp}_R({\bf x}) &= \int \frac{d^3 p}{\sqrt{(2\pi)^3}} \left[ {\widehat b}_R({\bf p}) w_1({\bf p})\ee^{\ii {\bf p} \cdot {\bf x}}  + {\widehat d}^\dagger_L({\bf p}) w_2(-{\bf p}) \ee^{ - \ii {\bf p} \cdot {\bf x}} \right] \,, \nonumber\\
{\widehat \vp}_L({\bf x}) &= \int \frac{d^3 p}{\sqrt{(2\pi)^3}} \left[ {\widehat b}_L({\bf p}) w_2({\bf p})\ee^{\ii {\bf p} \cdot {\bf x}}  + {\widehat d}^\dagger_R({\bf p}) w_1(-{\bf p}) \ee^{ - \ii {\bf p} \cdot {\bf x}} \right] \,,
\label{65}
\end{align}
where ${\widehat b}_c({\bf p})$ and ${\widehat d}_c({\bf p})$ respectively annihilate Weyl particles and anti-particles with momentum ${\bf p}$ and chirality $c$. They satisfy the usual anti-commutation relations.

We can also write
\be
{\widehat \vp}_R({\bf x}) = {\widehat b}_R({\bf x}) + {\widehat d}^\dagger_L({\bf x}) \,, \quad {\widehat \vp}_L({\bf x}) = {\widehat b}_L({\bf x}) + {\widehat d}^\dagger_R({\bf x}) \,. 
\label{66}
\en 
Using these expressions, the Hamiltonian ${\widehat H}$ can be written as ${\widehat H}_{W,1} +  {\widehat H}_{W,2}$, where ${\widehat H}_{W,2}$ corresponds to the term in ${\widehat H}$ which depends explicitly on the mass. We have that
\begin{align}
{\widehat H}_{W,1} &= \sum_{\alpha,\beta}\int d^3x \Big[ {\widehat b}^\dagger_{L,\alpha}({\bf x}) (\ii {\boldsymbol \sigma}_{\alpha \beta} \cdot {\boldsymbol \nabla}) {\widehat b}_{L,\beta}({\bf x}) + {\widehat d}_{R,\alpha}({\bf x}) (\ii {\boldsymbol \sigma}_{\alpha \beta} \cdot {\boldsymbol \nabla})  {\widehat d}^\dagger_{R,\beta}({\bf x}) \nonumber\\
& \qquad  - {\widehat b}^\dagger_{R,\alpha}({\bf x})  (\ii {\boldsymbol \sigma}_{\alpha \beta} \cdot {\boldsymbol \nabla}) {\widehat b}_{R,\beta}({\bf x}) -  {\widehat d}_{L,\alpha}({\bf x})  (\ii {\boldsymbol \sigma}_{\alpha \beta} \cdot {\boldsymbol \nabla}) {\widehat d}^\dagger_{L,\beta}({\bf x}) \Big] \,, \nonumber\\
{\widehat H}_{W,2} &= m \sum_{\alpha}\int d^3x \left(  {\widehat b}^\dagger_{L,\alpha}({\bf x}) {\widehat d}^\dagger_{L,\alpha}({\bf x}) + {\widehat b}^\dagger_{R,\alpha} {\widehat d}^\dagger_{R,\alpha}({\bf x}) +{\widehat d}_{L,\alpha}({\bf x}) {\widehat b}_{L,\alpha}({\bf x}) + {\widehat d}_{R,\alpha}({\bf x}) {\widehat b}_{R,\alpha}({\bf x}) \right) \,,
\label{68}
\end{align} 
where we have used certain properties of the spinors $w_i({\bf p)}$ given in~\cite{colin11}. Note that ${\widehat H}_{W,i} \neq {\widehat H}_{D,i}$ (${\widehat H}_{D,2}$ is not just the term in ${\widehat H}$ which explicitly depends on the mass). We further apply a normal ordering to the Hamiltonian. Note that the Hamiltonian $:{\widehat H}_{W,2}:$ does not commute with the total particle number. Pair creation and annihilation will be possible of a particle and an anti-particle with the same chirality.

Let us now introduce the states
\begin{multline}
|x,c,\alpha;n,{\bar n} \rangle_W =\\
 \frac{1}{\sqrt{n!{\bar n}!}}{\widehat b}^\dagger_{c_1,\alpha_1}({\bf x}_1) \dots {\widehat b}^\dagger_{c_n,\alpha_n}({\bf x}_n) {\widehat d}^\dagger_{c_{n+1}, \alpha_{n+1}}({\bf x}_{n+1}) \dots {\widehat d}^\dagger_{c_{n+{\bar n}},\alpha_{n+{\bar n}}}({\bf x}_{n+{\bar n}})|0\ra_W \,,
\label{69}
\end{multline}
where $\alpha=(\alpha_1,\dots,\alpha_{n + {\bar n}})$, with $\alpha_i=1,2$, and $|0\ra_W$ is the state that contains no left or right-handed chiral particles or anti-particles. Note that $|0\ra_W \neq |0\ra_D$; $|0\ra_D$ is an eigenstate of the Hamiltonian $:{\widehat H}:$ (with eigenvalue zero), whereas $|0\ra_W$ is not~\cite{colin11}. The overlaps ${}_W\la x,c,\alpha;n,{\bar n}| \Psi(t) \ra = \vp^{(n,{\bar n})}_{c,\alpha}(x,t)$ do not satisfy the many-particle equation \eqref{35}, since the Hamiltonian $:{\widehat H}:$ does not commute with the total particle number operator.

A possible choice of POVM is $P^{(n,{\bar n})}_{W,c}(dx) = \sum_\alpha |x,c,\alpha;n,{\bar n} \ra_W {}_W\la x,c,\alpha;n,{\bar n}| dx$. The velocity field corresponding to $:{\widehat H}_{W,1}:$ is the one given in \eqref{36}. The rates to jump from $(x,c)$, with $n$ particles and ${\bar n}$ anti-particles, to $(x',c')$, with $n'$ particles and ${\bar n}'$ anti-particles, are given by
\begin{multline}
t^{(n,{\bar n};n',{\bar n}')}_{cc'}(x,x',t) =   \frac{2}{\vp^{(n,{\bar n})\dagger}_{c}(x,t)\vp^{(n,{\bar n})}_{c}(x,t)} \\
\times \left[ {\textrm{Im}} \left(\sum_{\alpha,\alpha'}\vp^{(n',{\bar n}') *}_{c',\alpha'} (x',t){}_W\la x',c',\alpha';n',{\bar n}' |  :{\widehat H}_{W,2}: |x,c,\alpha;n,{\bar n} \rangle_W \vp^{(n,{\bar n})}_{c,\alpha}(x,t) \right) \right]^+ \,.
\label{70}
\end{multline}
We will not present the explicit expression for these jump rates. There will not be jumps solely of chirality. But there may be jumps corresponding to pair creation or annihilation of a particle and an anti-particle of the same chirality.

\subsubsection{Dirac sea}
Using ${\widehat {\widetilde b}}_{L}({\bf p}) = {\widehat d}^\dagger_R(-{\bf p})$ and ${\widehat {\widetilde b}}_{R}({\bf p}) = {\widehat d}^\dagger_L(-{\bf p})$, the field operators can also be written as
\begin{align}
{\widehat \vp}_R({\bf x}) &= \int \frac{d^3 p}{\sqrt{(2\pi)^3}} \left[ {\widehat b}_R({\bf p}) w_1({\bf p})\ee^{\ii {\bf p} \cdot {\bf x}}  + {\widehat {\widetilde b}}_{R}({\bf p}) w_2({\bf p}) \ee^{ \ii {\bf p} \cdot {\bf x}} \right] \,, \nonumber\\
{\widehat \vp}_L({\bf x}) &= \int \frac{d^3 p}{\sqrt{(2\pi)^3}} \left[ {\widehat b}_L({\bf p}) w_2({\bf p})\ee^{\ii {\bf p} \cdot {\bf x}}  + {\widehat {\widetilde b}}_{L}({\bf p}) w_1({\bf p}) \ee^{ \ii {\bf p} \cdot {\bf x}} \right] \,.
\label{81}
\end{align}
The operators ${\widehat {\widetilde b}}_R({\bf p})$ and ${\widehat {\widetilde b}}_L({\bf p})$ annihilate a particle with momentum ${\bf p}$, with negative energy $-|{\bf p}|$ and with respectively right- and left-handed chirality (and left- and right-handed helicity).  

The Hilbert space is spanned by the states
\be
|x,c,\alpha\rangle_W = \frac{1}{\sqrt{n!}} {\widehat \vp}^\dagger_{c_1,\alpha_1}({\bf x}_1)  \dots  {\widehat \vp}^\dagger_{c_n,\alpha_n}({\bf x}_n) |{\widetilde 0}\ra_W \,,
\label{82}
\en
where $|{\widetilde 0}\ra_W$ is the state that contains no Weyl particles of positive or negative energy, i.e., ${\widehat \vp}_{c,\alpha}({\bf x}) |{\widetilde 0}\ra_W = 0$, for all ${\bf x}$, $c=R,L$ and $\alpha_i=1,2$. Due to \eqref{51.1}, $|{\widetilde 0}\ra_W = |{\widetilde 0}\ra_D$ and $|x,c,\alpha\rangle_W = {\sqrt{2^n}}|x,c,\alpha\rangle_D$ for $\alpha =1,2$. The overlaps $\vp_{c,\alpha}(x,t) = {}_W\la x,c,\alpha | \Psi(t)\ra$ satisfy the many-particle wave equation \eqref{35}. 

Choosing the POVM $P_{W,c}(dx) = \sum_\alpha |x,c,\alpha \ra_W {}_W\la x,c,\alpha| dx = P_{D,c}(dx)$, the pilot-wave model is the same as the one obtained in section \ref{dirac-diracsea} starting from the Dirac sea picture for massive Dirac particles. This equivalence still holds in the case of interactions.

Starting from the position POVM $\sum_c P_{W,c}(dx) = P_D (dx)$ the deterministic pilot-wave model of~\cite{colin07} is obtained. So in this approach, particles are not distinguished by chirality. One could also say that in this model positions are introduced for massless Dirac particles~\cite{colin11} (because the operators ${\widehat b}_c({\bf p})$ can also be interpreted as the annihilation operator of a positive energy massless Dirac particle with helicity $c$ and similarly for the operators ${\widehat {\widetilde b}}_c({\bf p})$).

\subsection{Summary}\label{differentmodels}
We have discussed a number of possible pilot-wave approaches for the quantized Dirac field. In the context of the standard model, where the particles are fundamentally massless, one should of course introduce positions for the massless particles. We have discussed four ways of doing this. One can choose to adopt the particle--anti-particle picture or the Dirac sea picture. One can further choose whether or not to distinguish particles of different chirality. In the particle--anti-particle picture, there is the possibility of pair creation, even in the free case, whether or not one chooses to distinguish particles of different chirality. In the Dirac sea picture, there are no jumps of position. If one introduces particles with different chirality then there might be jumps of chirality, otherwise the model is deterministic (even in the case of fermion number preserving interactions).

\section{Non-relativistic limit}\label{nonrelativistic}
In this section, we derive the non-relativistic limit of the zig-zag pilot-wave model for the case of a single particle. 

\subsection{Non-relativistic limit of the Dirac equation}
We first consider the non-relativistic limit of the Dirac equation, following the usual account as found in e.g.~\cite{hill38,pauli80}. Starting from the Dirac-Pauli representation and the ansatz
\begin{equation}
\psi = \left( \begin{array}{c} {\widetilde \varphi} \\{\widetilde \chi}  \end{array} \right) =  \ee^{-\ii m t} \left( \begin{array}{c} \varphi \\ \chi  \end{array} \right) \,,
\label{100}
\end{equation}
and assuming $|\ii \pa_t \chi - e V \chi| \ll m |\chi|$, $\chi$ can be expanded as 
\begin{equation}
\chi = \chi_1 + \chi_3 + \dots \,,
\label{101}
\end{equation}
where 
\begin{equation}
\chi_1 = - \frac{\ii}{2m} {\boldsymbol \sigma} \cdot {\bf D}\varphi \,, \qquad \chi_3 = - \frac{\ii}{8m^3} ({\boldsymbol \sigma} \cdot {\bf D})^3 \varphi - \frac{\ii e}{4m^2} {\boldsymbol \sigma} \cdot {\bf E} \varphi  \,,
\label{102}
\end{equation}
with ${\bf D} = {\boldsymbol \nabla} - \ii e{\bf A}$. ${\bf E}= - \pa_t {\bf A} - {\boldsymbol \nabla}V$ and ${\bf B} = {\boldsymbol \nabla} \times {\bf A}$ are respectively the electric and magnetic field. As such, $\chi$ is the small component relative to $\varphi$, with $|\chi_1|$ and $|\chi_3|$ respectively of the order $  (p/m)|\varphi|$ and $( p^3/m^3) |\varphi|$. The dots in \eqref{101} denote the higher order terms. 

To lowest order, elimination of $\chi$ in favor of $\vp$ yields the Pauli equation:
\begin{equation}
\ii \pa_t \vp = - \frac{1}{2m} D^2 \vp - \frac{e}{2m} {\bf B} \cdot {\boldsymbol \sigma} \vp  + e V \vp \,.
\label{103}
\end{equation}
The next order term in the Hamiltonian is of the order $p^4/m^3$ and only arises when $\chi_3$ is taken into account. Those higher order terms will not be considered. 

The expansion \eqref{101} can be used to expand the current 
\begin{equation}
j^\mu_c = j^\mu_{c,0} + j^\mu_{c,1} + j^\mu_{c,2} + \dots \,,
\label{104}
\end{equation}
where $c=R,L$ and the terms $j^\mu_{c,i}$ are of the order $p^i/m^i$. This yields
\begin{align}
j^0_{c,0} & = \frac{1}{2}\vp^\dagger \vp \,, \nonumber\\
j^0_{c,1} & =  s(c) \frac{1}{2m} {\textrm{Im}} \left(\vp^\dagger {\boldsymbol \sigma}\cdot {\bf D} \vp \right)  \,, \nonumber\\
j^0_{c,2} & = \frac{1}{8 m^2} \left( {\boldsymbol \sigma}\cdot {\bf D} \vp \right)^\dagger \left( {\boldsymbol \sigma}\cdot {\bf D} \vp \right) 
=\frac{1}{8 m^2} \left[  \left(  {\bf D} \vp \right)^\dagger \cdot \left( {\bf D} \vp - \ii  {\boldsymbol \sigma} \times  {\bf D} \vp  \right)       \right]\,,
\label{105}
\end{align}
and 
\begin{align}
{\bf j}_{c,0} & = s(c) \frac{1}{2}\vp^\dagger {\boldsymbol \sigma} \vp  \,, \nonumber\\
{\bf j}_{c,1} & = \frac{1}{2m} {\textrm{Im}} \left(\vp^\dagger {\bf D} \vp \right) + \frac{1}{4m} {\boldsymbol \nabla} \times \left(\vp^\dagger {\boldsymbol \sigma} \vp \right)  \,, \nonumber\\
{\bf j}_{c,2} & = s(c)  \frac{1}{8 m^2} \left( {\boldsymbol \sigma}\cdot {\bf D} \vp \right)^\dagger {\boldsymbol \sigma}\left( {\boldsymbol \sigma}\cdot {\bf D} \vp \right)   \nonumber\\
&=  s(c) \frac{1}{8 m^2}\left[  \left( {\boldsymbol \sigma}\cdot {\bf D} \vp \right)^\dagger {\bf D} \vp + \left(  {\bf D} \vp \right)^\dagger {\boldsymbol \sigma}\cdot {\bf D} \vp - \left(  {\bf D} \vp \right)^\dagger\cdot {\bf D} {\boldsymbol \sigma} \vp - \ii \left(  {\bf D} \vp \right)^\dagger \times  {\bf D} \vp  \right] \,,
\label{106}
\end{align}
with $s(c)$ as defined in \eqref{35.1}. Similarly, $F$, which is given in \eqref{15}, can be expanded into terms $F_i$ which are of the order $p^{i}/m^{i-1}$. The lowest order terms read 
\begin{align}
F_0 &= 0\,, \nonumber\\
F_1 &=  \frac{1}{2}{\boldsymbol \nabla}  \cdot  \left( \vp^\dagger {\boldsymbol \sigma} \vp \right) \,,\nonumber\\
F_2 &= 0 \,,\nonumber \\
F_3 &= \frac{1}{4 m^2} {\textrm{Re}}\left(\vp^\dagger ({\boldsymbol \sigma}\cdot {\bf D})^3  \vp \right)  + \frac{e}{2m} {\bf E} \cdot \left( \vp^\dagger {\boldsymbol \sigma} \vp \right) \nonumber \\
&= \frac{1}{4 m^2} \left( {\textrm{Re}} \left[ \vp^\dagger {\bf D} \cdot ({\boldsymbol \sigma}\cdot {\bf D})  {\bf D} \vp \right]  + \frac{e}{2} {\bf B}\cdot {\boldsymbol \nabla}(\vp^\dagger \vp)  \right) + \frac{e}{2m} {\bf E} \cdot \left( \vp^\dagger {\boldsymbol \sigma} \vp \right)\,.
\label{107}
\end{align}

Assuming that $\vp$ satisfies the Pauli equation \eqref{103}, we have the following identities:
\begin{align}
 {\boldsymbol \nabla} \cdot {\bf j}_{c,0} &= s(c)F_1 \,, \label{108}\\
 \pa_t j^0_{c,0}  + {\boldsymbol \nabla} \cdot {\bf j}_{c,1} &= 0 \,, \label{109}\\
 \pa_t j^0_{c,1}  + {\boldsymbol \nabla} \cdot {\bf j}_{c,2} &= s(c)F_3 \,.
\label{110}
\end{align}
This can be verified by direct calculation. Alternatively, it can be seen by considering the expansion of the identity $\pa_\mu j^\mu_c = s(c)F$ using \eqref{104}-\eqref{107}. The equations \eqref{108}-\eqref{110} correspond to the terms in this expansion of the order $p^{i+1}/m^i$, with $i=0,1,2$. Namely, the terms ${\boldsymbol \nabla} \cdot {\bf j}_{c,i}$ are of order $p^{i+1}/m^i$ and, assuming the Pauli equation to evaluate the time derivatives, the terms $\pa_t j^0_{c,i}$ are of order $p^{i+2}/m^{i+1}$. Higher order corrections to the Pauli equation, which are of the order $p^4/m^3$, only show up in higher order terms in $\pa_\mu j^\mu_c = s(c)F$, which are not considered here.

We also have the identities:
\be
\left( j^0_{c,0} \right)^2 = |{\bf j}_{c,0}|^2 \,, \quad j^0_{c,0} j^0_{c,1} = {\bf j}_{c,0} \cdot {\bf j}_{c,1} \,, \quad \left( j^0_{c,1} \right)^2 + 2 j^0_{c,0} j^0_{c,2} =  |{\bf j}_{c,1}|^2 + 2 {\bf j}_{c,0} \cdot {\bf j}_{c,2} \,.
\label{110.1}
\en 
As before, they can be found by considering the expansion of the identities $j^\mu_cj_{c\mu}  =0$ using \eqref{104}-\eqref{107}. 

\subsection{Non-relativistic limit of Bohm's approach}
Let us now consider the non-relativistic limit of Bohm's approach to the Dirac theory (which was derived before in~\cite{bohm93}). In this approach, the velocity field of the particle is given by ${\bf v}_D = {\bf j}_{D} / j^0_{D}$, with $j^\mu_D$ the Dirac current defined in \eqref{2}. Since $j^\mu_D  = j^\mu_R + j^\mu_L$, we have from \eqref{105} and \eqref{106} that
\begin{equation}
j^0_{D,0} = \vp^\dagger \vp  \,, \qquad j^0_{D,1} = 0  
\label{111}
\end{equation}
and
\begin{equation}
{\bf j}_{D,0} = 0 \,, \qquad  {\bf j}_{D,1} = \frac{1}{m} {\textrm{Im}} \left(\vp^\dagger {\bf D} \vp \right) + \frac{1}{2m} {\boldsymbol \nabla} \times \left(\vp^\dagger {\boldsymbol \sigma} \vp \right) \,, \qquad {\bf j}_{D,2} = 0 \,.
\label{112}
\end{equation}
Hence, up to order $p^2/m^2$, the Dirac current ${\bf j}_{D}$ is given by ${\bf j}_{D,1}$. This forms a current for the Pauli theory and we will denote it by ${\bf j}_P$. Namely, assuming the Pauli equation, it satisfies the conservation equation
\begin{equation}
\pa_t ( \vp^\dagger \vp ) + {\boldsymbol \nabla} \cdot {\bf j}_P =0 
\label{113}
\end{equation}
(which is just a rewriting of \eqref{109}).

Up to order $p^2/m^2$, the Dirac velocity field ${\bf v}_D$ reduces to the Pauli velocity field 
\be
{\bf v}_P= \frac{{\bf j}_P}{ \vp^\dagger \vp} =  \frac{1}{m\vp^\dagger \vp} {\textrm{Im}} \left(\vp^\dagger {\bf D} \vp \right) + \frac{1}{2m\vp^\dagger \vp} {\boldsymbol \nabla} \times \left(\vp^\dagger {\boldsymbol \sigma} \vp \right)  \,.
\label{114}
\en
The Pauli velocity field can be used to yield a pilot-wave approach to the Pauli equation. Namely, in virtue of the continuity equation \eqref{113}, the corresponding particle dynamics guarantees the equivariance of the Pauli distribution $\vp^\dagger \vp$.

\subsection{Zig-zag pilot-wave approaches for the Pauli theory}\label{zzpauli}
Before passing to the non-relativistic limit of the zig-zag pilot-wave approach for the Dirac theory, it is worth considering the following two pilot-wave approaches for the Pauli theory. They are suggested by considering the equations \eqref{7} respectively up to order $p^3/m^2$ and $p^2/m$ (which correspond to combinations of the identities \eqref{108}-\eqref{110}). In the next section, we will compare these approaches to the non-relativistic limit of the zig-zag pilot-wave approach for the Dirac theory.
 
In the first case, a combination of \eqref{108}-\eqref{110} yields
\be
\pa_t \left( j^0_{c,0} + j^0_{c,1} \right) + {\boldsymbol \nabla} \cdot \left( {\bf j}_{c,0} + {\bf j}_{c,1} + {\bf j}_{c,2}\right) =  s(c)(F_1 + F_3) \,,
\label{115}
\en
where $c=R,L$ and where the Pauli equation is assumed. This suggests a pilot-wave approach with velocity fields 
\be
{\widetilde {\bf v}}_c = \frac{{\bf j}_{c,0} + {\bf j}_{c,1} + {\bf j}_{c,2}}{j^0_{c,0} + j^0_{c,1}} 
\label{116}
\en
and jump rates
\begin{equation}
{\widetilde t}_{c,\pi c} = \frac{\left[-s(c)(F_1 + F_3)\right]^+}{j^0_{c,0} + j^0_{c,1}} 
\label{117}
\end{equation}
for the velocity field to respectively jump from ${\widetilde {\bf v}}_c$ to ${\widetilde {\bf v}}_{\pi c}$. The equivariant equilibrium distribution is given by $j^0_{c,0} + j^0_{c,1}$ (which is positive under the assumption that $j^0_{c,0} \gg |j^0_{c,1}|$). The resulting position density is $j^0_{R,0} + j^0_{L,0} =  \vp^\dagger \vp$, so that this approach reproduces the usual predictions of the Pauli theory.

In the second case, a combination of \eqref{108} and \eqref{109} yields
\be
\pa_t   j^0_{c,0}  + {\boldsymbol \nabla} \cdot \left( {\bf j}_{c,0} + {\bf j}_{c,1} \right) =  s(c) F_1 \,.
\label{119}
\en
This suggests a pilot-wave approach with velocity fields 
\be
{\bar {\bf v}}_c = \frac{{\bf j}_{c,0} + {\bf j}_{c,1}}{j^0_{c,0}} = s(c) \frac{ \vp^\dagger {\boldsymbol \sigma} \vp} {\vp^\dagger \vp} + \frac{{\bf j}_P}{\vp^\dagger \vp} = s(c) {\bf s} +{\bf v}_P 
\label{120}
\en
and jump rates
\begin{equation}
{\bar t}_{c,\pi c} = \frac{\left[-s(c)F_1\right]^+}{j^0_{c,0}} =  \frac{ \left[-s(c) {\boldsymbol \nabla} \cdot \left( \vp^\dagger {\boldsymbol \sigma} \vp \right)\right]^+ } {\vp^\dagger \vp} \,,
\label{121}
\end{equation}
where the vector ${\bf s}=\vp^\dagger {\boldsymbol \sigma} \vp/\vp^\dagger \vp$ is the normalized spin polarization vector and ${\bf v}_P$ is the usual Pauli velocity field. The equivariant equilibrium distribution is given by $j^0_{c,0}=\vp^\dagger \vp/2$ and again yields $\vp^\dagger \vp$ as position density. So also this approach reproduces the usual predictions of the Pauli theory.

\subsection{Non-relativistic limit of the zig-zag approach for the Dirac theory}
Turning to the zig-zag approach for the Dirac theory, we have that, up to terms of the order $p^2/m^2$, the velocity fields read
\begin{equation}
{\bf v}_c = \frac{1}{j^0_{c,0}} \left(  {\bf j}_{c,0} + {\bf j}_{c,1} + {\bf j}_{c,2}  - \frac{j^0_{c,1}}{j^0_{c,0}} \left({\bf j}_{c,0} + {\bf j}_{c,1} \right) + \left[ \left( \frac{j^0_{c,1}}{j^0_{c,0}} \right)^2  -  \frac{j^0_{c,2}}{j^0_{c,0}} \right] {\bf j}_{c,0} \right) + {\mathcal{O}}\left( \frac{p^3}{m^3} \right)
\label{123}
\end{equation}
and up to terms of the order $p^3/m^2$, the jump rates read
\begin{equation}
t_{c,\pi c} =  \frac{1}{j^0_{c,0}}  \left( \left[ -s(c) (F_1 + F_3)\right]^+ + \left[ \left( \frac{j^0_{c,1}}{j^0_{c,0}} \right)^2 - \frac{j^0_{c,1} + j^0_{c,2}}{j^0_{c,0}} \right] (-s(c) F_1)^+ \right)+ {\mathcal{O}}\left( \frac{p^4}{m^3} \right)\,.
\label{124}
\end{equation}

Let us now compare this to the first model for the Pauli theory considered in the previous section. When expanding ${\widetilde {\bf v}}_c$ to the same order, we find the same expression, up to the term containing $j^0_{c,2}$ (which is of the order $p^2/m^2$). Similarly, when expanding ${\widetilde t}_{c,\pi c}$ to the order $p^3/m^2$, it is equal to the above expression, up to the the term containing $j^0_{c,2}$. So by considering the velocity fields and the jump rates up to the present order, we do not obtain exactly a pilot-wave approach to the Pauli equation. Namely, the dynamics does not ensure the equivariance of the distributions $j^0_{c,0} + j^0_{c,1}$ exactly, but only up to terms of order $p^2/m$. As such, the position distribution $\vp^\dagger \vp$ is only approximately preserved.

Note that the extra term in the expansion of ${\bf v}_c$ compared to ${\widetilde {\bf v}}_c$ follows from the fact that $|{\bf v}_c| = 1$. The expression in the right hand side of \eqref{123} has norm one, up to order  $p^2/m^2$ (which can be verified explicitly using the identities \eqref{110.1}), whereas ${\widetilde {\bf v}}_c$ does not.

A similar conclusion holds if we consider the non-relativistic limit up to a lower order. Namely, up to terms of the order $p/m$, we have
\begin{align}
{\bf v}_c &= \frac{1}{j^0_{c,0}} \left(  {\bf j}_{c,0} + {\bf j}_{c,1}   - \frac{j^0_{c,1}}{j^0_{c,0}} {\bf j}_{c,0}  \right) + {\mathcal{O}}\left( \frac{p^2}{m^2} \right) \nonumber\\
&=  {\bar {\bf v}}_c  + {\bf v}_A  + {\mathcal{O}}\left( \frac{p^2}{m^2} \right)\nonumber\\
&=  s(c) {\bf s} + {\bf v}_P + {\bf v}_A + {\mathcal{O}}\left( \frac{p^2}{m^2} \right) \,,
\label{125}
\end{align}
where \eqref{120} was used and
\be
{\bf v}_A = - \frac{j^0_{c,1}}{ j^0_{c,0} } \frac{{\bf j}_{c,0}}{j^0_{c,0}} = -  \frac{1}{m} \frac{{\textrm{Im}\left(\vp^\dagger {\boldsymbol \sigma}\cdot {\bf D} \vp \right) }}{\vp^\dagger \vp} \frac{ \vp^\dagger {\boldsymbol \sigma} \vp} {\vp^\dagger \vp} = -  \frac{1}{m} \frac{{\textrm{Im}\left(\vp^\dagger {\boldsymbol \sigma}\cdot {\bf D} \vp \right) }}{\vp^\dagger \vp} {\bf s} \,.
\en
The jump rates reduce to 
\begin{align}
t_{c,\pi c} &= \frac{ (-s(c) F_1)^+ }{j^0_{c,0}}  \left(1 - \frac{j^0_{c,1} }{j^0_{c,0}}\right) + {\mathcal{O}}\left( \frac{p^3}{m^2} \right) \nonumber\\
&= {\bar t}_{c,\pi c} \left( 1 -s(c) \frac{1}{m} \frac{{\textrm{Im}\left(\vp^\dagger {\boldsymbol \sigma}\cdot {\bf D} \vp \right) }}{\vp^\dagger \vp} \right) + {\mathcal{O}}\left( \frac{p^3}{m^2} \right) \nonumber\\
&= {\bar t}_{c,\pi c} \left( 1 + s(c) {\bf s} \cdot   {\bf v}_A   \right) + {\mathcal{O}}\left( \frac{p^3}{m^2} \right)\,.
\label{126}
\end{align}

The additional term ${\bf v}_A$ compared to ${\bar {\bf v}}_c$ is of order $p/m$ and ensures that $s(c) {\bf s} + {\bf v}_P + {\bf v}_A$ has norm one, up to order $p/m$ (which can be verified explicitly using \eqref{110.1}). Note that we can also write ${\bf v}_A = - \left({\bf v}_P \cdot {\bf s} \right){\bf s} $ (using \eqref{110.1}). This implies that the additional terms in the velocity fields and the jump rates only vanish when the usual Pauli velocity field is orthogonal to the spin vector.

Note that the dominating term in the velocity field ${\bf v}_c$ is ${\bf j}_{c,0}/j^0_{c,0} = s(c) {\bf s}$, which has norm one. So up to lowest order of approximation, particles move back and forth at the speed of light, along the direction of the spin vector.

Again we do not obtain exactly a pilot-wave approach to the Pauli equation. Namely the dynamics does not exactly preserve the position distribution $\vp^\dagger \vp$.

\subsection{Spin eigenstate}
For the special case of vanishing electromagnetic potentials and a Pauli spinor which is a spin eigenstate, i.e., $\vp({\bf x},t) = \psi({\bf x},t) \xi$, where $\psi$ is a scalar function and $\xi^\dagger \xi= 1$, the Pauli equation reduces to the non-relativistic Schr\"odinger equation for $\psi$. 

Writing $\psi = |\psi|\ee^{\ii S}$ and using ${\bf s} = \xi^\dagger {\boldsymbol \sigma} \xi$, the lowest order terms in the expansion of the current and the quantity $F$ obtain the following form:
\begin{align}
j^0_{c,0} & = \frac{1}{2}|\psi|^2 \,,\\
j^0_{c,1} & = s(c) \frac{1}{2m}|\psi|^2 {\bf s} \cdot  {\boldsymbol \nabla} S \,,\\
j^0_{c,2} & = \frac{1}{8 m^2} \left[  |{\boldsymbol \nabla} \psi|^2  +  \ii  {\bf s} \cdot \left(  {\boldsymbol \nabla} \psi^*  \times  {\boldsymbol \nabla} \psi  \right)   \right]   \,,\\
{\bf j}_{c,0} & = s(c) \frac{1}{2} {\bf s} |\psi|^2\,,\\
{\bf j}_{c,1} & = \frac{1}{2m} |\psi|^2 {\boldsymbol \nabla} S  + \frac{1}{4m} {\boldsymbol \nabla} \times \left( {\bf s} |\psi|^2   \right) \,,\\
{\bf j}_{c,2} & = s(c) \frac{1}{8 m^2}  \left[ {\bf s} \cdot  {\boldsymbol \nabla} \psi^*  {\boldsymbol \nabla} \psi + {\boldsymbol \nabla} \psi^* {\bf s} \cdot  {\boldsymbol \nabla} \psi - {\bf s} |{\boldsymbol \nabla} \psi|^2 - \ii {\boldsymbol \nabla} \psi^* \times {\boldsymbol \nabla} \psi \right]\,,
\label{128}
\end{align}
and
\be
F_0 = 0\,, \quad F_1 = \frac{1}{2}{\bf s} \cdot  {\boldsymbol \nabla} |\psi|^2 \,,\quad F_2 =0\,, \quad F_3 =  \frac{1}{4 m^2}  {\textrm{Re}} \left( \psi^* {\bf s} \cdot  {\boldsymbol \nabla} \nabla^2 \psi   \right) \,.
\label{129}
\en
These expressions can be used to obtain the form of the zig-zag pilot-wave models for the Pauli theory considered in section \ref{zzpauli} and the non-relativistic limit of the model for the Dirac theory. 

Let us first consider the zig-zag model for the Pauli theory defined by \eqref{120} and \eqref{121}. The velocity fields reduce to
\begin{equation}
{\bar {\bf v}}_c = s(c) {\bf s} + \frac{1}{m} {\boldsymbol \nabla} S + \frac{1}{2m} {\boldsymbol \nabla} \times \left({\bf s} \ln |\psi|^2   \right) \,, 
\label{130}
\end{equation}
and the jump rates to
\begin{equation}
{\bar t}_{c,\pi c} = \left[ - s(c) {\bf s} \cdot  {\boldsymbol \nabla} \ln |\psi|^2 \right]^+  \,.
\label{131}
\end{equation}
So the velocity field is given by the usual velocity for the non-relativistic Schr\"odinger theory, namely ${\boldsymbol \nabla} S/m$, plus a curl term (which needs to be added to the usual velocity field when one starts from the usual pilot-wave approach to the Dirac theory~\cite{holland03a}), and plus or minus the spin vector. The jump rates are smaller when the spatial variation of density $|\psi|^2$ is slower. 

In the case of the zig-zag pilot-wave model for the Dirac theory, the velocity fields reduce to
\begin{equation}
{\bf v}_c = {\bar {\bf v}}_c  - \frac{1}{m}  {\bf s } ({\bf s} \cdot  {\boldsymbol \nabla} S) + {\mathcal{O}}\left( \frac{p^2}{m^2} \right)  \label{132}
\end{equation}
and the jump rates to
\begin{equation}
t_{c,\pi c} = {\bar t}_{c,\pi c} \left( 1 - s(c) \frac{1}{m} {\bf s} \cdot  {\boldsymbol \nabla} S \right) + {\mathcal{O}}\left( \frac{p^3}{m^2} \right) \\,.
\label{133}
\end{equation}
The term in the velocity field ${\bf v}_c$ that is additional to ${\bar {\bf v}}_c$ corresponds to minus the component of the usual \dbb\ velocity field ${\boldsymbol \nabla} S/m$ for the non-relativistic Schr\"odinger theory along the direction of the spin vector. 

Similarly as in the case of a general Pauli spinor, the expressions \eqref{130} and \eqref{131} determine a pilot-wave model for the non-relativistic Schr\"odinger theory because they preserve the density $|\psi|^2$, while the expressions \eqref{132} and \eqref{133} do not.

\section{Variations on a theme}\label{variations}
We have explored the zig-zag pilot-wave model for the Dirac theory. This model forms a stochastic alternative to Bohm's original deterministic approach. For a single particle, the stochasticity arises from the fact that the velocity field might jump between two alternatives, respectively determined by the left- and right-handed chiral component of the Dirac spinor. It is clear that still other pilot-wave models could be obtained by starting from a different decomposition of the Dirac spinor. For example, one could decompose it into two Majorana spinors, or one could decompose it into the two spin components along a certain direction. Since such approaches seem rather unnatural, we will not pursue them further. We will just illustrate certain aspects of such alternative models in the context of the Pauli theory and the non-relativistic Schr\"odinger theory. Again, there seems to be no reason to consider these models as serious alternatives to the usual models. 

\subsection{An alternative pilot-wave model for the Pauli theory}
Reconsidering the Pauli equation \eqref{103}, a different pilot-wave model than the usual one can be obtained as follows. Writing
\begin{equation}
\vp = \left( \begin{array}{c} \vp_1 \\ \vp_2  \end{array} \right) \,,
\label{150}
\end{equation}
with $\vp_1$ and $\vp_2$ respectively the spin-up and spin-down component along the $z$-axis, the Pauli equation implies the following identities:
\begin{equation}
\pa_t |\vp_1|^2 + {\boldsymbol \nabla} \cdot {\bf j}_1 = I \,, \qquad \pa_t |\vp_2|^2 + {\boldsymbol \nabla} \cdot {\bf j}_2 = -I \,,
\label{151}
\end{equation}
where
\begin{equation}
{\bf j}_i= \frac{1}{m} {\textrm{Im}}  (\vp^*_i {\bf D} \vp_i ) \,, \quad i=1,2 \,,
\label{152} 
\end{equation}
and
\begin{equation}
I = \frac{e}{m} {\textrm{Im}} \left( (B_1 + \ii B_2) \vp^*_2 \vp_1 \right) \,.
\label{153} 
\end{equation}
This suggests a pilot-wave model, in which a point-particle moves along continuous trajectories, with a velocity field that jumps between ${\bf v}_1 = {\bf j}_1/|\vp_1|^2 $ and  ${\bf v}_2 = {\bf j}_2/|\vp_2|^2$, with jump rates $t_{21} = I^+ / |\vp_2|^2$ and $t_{12} = (-I)^+ / |\vp_1|^2$ to jump from ${\bf v}_2$ to ${\bf v}_1$ and vice versa. As such, the particle is either guided by the spin-up or spin-down component. As a distribution on velocity phase space, the equivariant equilibrium distribution is given by
\be
\rho({\bf x},{\bf v},t) =   |\vp_1({\bf x},t)|^2 \delta({\bf v} -{\bf v}_1 ({\bf x},t)) + |\vp_2({\bf x},t)|^2   \delta({\bf v} - {\bf v}_2({\bf x},t)) \,.
\label{153.1} 
\en
The corresponding position distribution is $\vp^\dagger \vp = |\vp_1|^2 + |\vp_2|^2$, so that the model is empirically equivalent to the usual model for the Pauli theory.

Note that in the special case of vanishing electromagnetic potentials and a spin eigenstate $\vp({\bf x},t) = \psi({\bf x},t) \xi$, this model reduces to the usual non-relativistic pilot-wave theory of \db\ and Bohm for spinless particles. Namely, in that case, the jump rates are zero and the velocity field is given by ${\boldsymbol \nabla} S/m$, where $\psi = |\psi|\ee^{\ii S}$.

Since we started from a decomposition of $\vp$ into its spin components along the $z$-axis, the obtained pilot-wave model is not rotationally invariant. In~\cite{deangelis82} a similar theory was proposed in the context of Nelson's stochastic theory. There it was noted that, in the case of a homogeneous magnetic field, rotational invariance could be achieved by decomposing the spinor into its spin components along the direction of the magnetic field. For a non-homogeneous magnetic field one could consider a decomposition of $\vp$ that varies spatially as well as temporally. However, the magnetic field could be zero in certain regions of space, leaving such a decomposition ill-defined. A better possibility would be to consider a decomposition along the direction of the spin vector ${\bf s}= \vp^\dagger {\boldsymbol \sigma} \vp/\vp^\dagger \vp$ which is non-zero when $\vp$ is non-zero. However, since such a construction appears rather unnatural and complicated, we will not pursue it further.

\subsection{An alternative pilot-wave model for the non-relativistic Schr\"odinger theory}
Consider now the non-relativistic Schr\"odinger equation
\begin{equation}
\ii \pa_t \psi = - \frac{1}{2m} D^2 \psi + V\psi \,.
\label{154} 
\end{equation}
We can write $\psi = \psi_1 + \ii \psi_2$, where $\psi_1$ and $\psi_2$ are real. This leads to 
\begin{equation}
\pa_t \psi_1^2 + {\boldsymbol \nabla} \cdot {\bf j}_1 = I \,, \qquad \pa_t \psi_2^2 + {\boldsymbol \nabla} \cdot {\bf j}_2 = -I \,,
\label{155}
\end{equation}
where
\begin{equation}
{\bf j}_1 = \frac{1}{m} \psi_1 \left( {\boldsymbol \nabla} \psi_2 - e {\bf A} \psi_1 \right) \,, \quad {\bf j}_2 = - \frac{1}{m} \psi_2  \left( {\boldsymbol \nabla} \psi_1 + e{\bf A} \psi_2 \right)\,,
\label{156} 
\end{equation}
and
\begin{equation}
I = \frac{1}{m} {\boldsymbol \nabla} \psi_1  \cdot {\boldsymbol \nabla} \psi_2  + \frac{e}{m} |{\bf A}|^2 \psi_1 \psi_2    + 2V \psi_1 \psi_2  \,. 
\label{157} 
\end{equation}
This suggests a pilot-wave model where the velocity jumps between ${\bf v}_1 = {\bf j}_1/\psi^2_1 $ and ${\bf v}_2 = {\bf j}_2/\psi_2^2$, with jump rates $t_{21} = I^+ / \psi_2^2$ and $t_{12} = (-I)^+ / \psi_1^2$. Note that in this case the velocity fields depend on both components of the wave function. The equilibrium distribution is given by
\be
\rho({\bf x},{\bf v},t) =   \psi_1({\bf x},t)^2 \delta({\bf v} -{\bf v}_1 ({\bf x},t)) + \psi_2({\bf x},t)^2   \delta({\bf v} - {\bf v}_2({\bf x},t)) \,,
\label{158} 
\en
with corresponding position distribution $ |\psi|^2$.  

The obtained model is not gauge invariant. Namely, a gauge transformation $\psi \to \ee^{\ii e \theta} \psi$, ${\bf A} \to {\bf A}- {\boldsymbol \nabla}\theta$ will yield different trajectories. Nevertheless, in equilibrium, the violation of gauge invariance can not be detected. In the usual pilot-wave approach of \db\ and Bohm, particles move with a velocity ${\textrm{Im}}( {\boldsymbol \nabla} \psi/\psi) / m  = ({\bf j}_1 + {\bf j}_2)/ |\psi|^2$, which is invariant under gauge transformations. 

The decomposition of the Dirac spinor into two Majorana spinors (which corresponds to a splitting into real and imaginary part in the Majorana representation of the Dirac matrices) is similarly not invariant under $U(1)$ gauge transformations. Hence, a pilot-wave model in the spirit of the zig-zag model, starting from such a decomposition, would also not be gauge invariant.

\section{Acknowledgments}
It is a pleasure to thank Samuel Colin and Christian Maes for valuable discussions. This work was supported by the FWO project G.0647.11.

\end{document}